\documentclass[%
 reprint,
 amsmath,amssymb,
 aps,
prl,
]{revtex4-1}

\usepackage{graphicx}
\usepackage{dcolumn}
\usepackage{bm}
\usepackage{color}
\usepackage{floatrow}
\usepackage[label font=bf,labelformat=simple]{subfig}
\usepackage{caption}
\begin{document}

\preprint{APS/123-QED}

\title{Noise-Resilient Quantum Dynamics Using Symmetry-Preserving Ansatzes}
  
\author{Matthew Otten}
\email[Correspondence: ]{otten@anl.gov}
\author{Cristian L. Cortes}
\author{Stephen K. Gray}
 \affiliation{%
 Center for Nanoscale Materials, Argonne National Laboratory, Lemont, Illinois, 60439
 }%

\date{\today}

\begin{abstract}
  We describe and demonstrate a method for the computation of quantum
  dynamics on small, noisy universal quantum computers.  This method relies on the
idea of `restarting' the dynamics; at least one approximate time step is taken
on the quantum computer and then a parameterized quantum circuit ansatz is
optimized to produce a state that well approximates the time-stepped results.
The simulation is then restarted from the optimized state. By
encoding knowledge of the form of the solution in the ansatz, such as ensuring
that the ansatz has the appropriate symmetries of the Hamiltonian, the optimized
ansatz can recover from the effects of decoherence.
This allows for the
quantum dynamics 
to proceed far beyond the standard gate depth limits of the underlying hardware,
albeit incurring some error from the optimization, the quality of the ansatz,
and the typical time step error.
We demonstrate this methods on the Aubry-Andr\'e model with interactions at
half-filling, which  
shows interesting many-body localization effects in the long time
limit. Our 
method is capable of performing high-fidelity Hamiltonian simulation hundred of
time steps longer than the standard Trotter
approach. These results demonstrate a path towards using small, lossy devices to
calculate quantum dynamics. 
\end{abstract}

\pacs{Valid PACS appear here}
\maketitle
\textit{Introduction.}
Quantum dynamics was one of the first quantum computing applications
envisioned~\cite{feynman1982simulating} 
and still remains one of the most promising. Though some quantum devices can be
built to specifically simulate 
the dynamics of a single or small class of quantum systems
exactly~\cite{buluta2009quantum,debnath2018observation}, universal,
gate-based
quantum computers require some 
approximate time-stepping scheme. One of the most well-studied is the
the Trotter or more generally Trotter-Suzuki
decomposition~\cite{berry2007efficient}, which decomposes the propagator
for an arbitrary
Hamiltonian into a sequence of gates which can be computed on universal quantum 
devices. Reaching long times requires gate depths well beyond those achievable
on near-term quantum devices~\cite{martinez2016real}.
 Hybrid quantum-classical methods,
especially in quantum
chemistry~\cite{omalley-prx-2016,peruzzo-ncomms-2014,kandala-nature-2017} 
and quantum machine learning~\cite{havlivcek2019supervised,biamonte2017quantum},
can alleviate the need for high gate depth by 
using variational methods with much shorter circuits.
There have been proposals for using variational methods to do quantum dynamics
on quantum computers,
relying on variational principles~\cite{li2017efficient} or subspace
expansion~\cite{heya2019subspace}, which can provide the 
dynamics of a variational wavefunction using short circuits, but they require an
extra, all-to-all connected qubit which computes the derivatives with respect to
the variational parameters~\cite{yuan2019theory} or are limited to the dynamics
of the ground and low-lying excited states~\cite{heya2019subspace}. Another
recent method uses variational diagonalization to allow for variational fast
forwarding of the dynamics~\cite{cirstoiu2019variational}.

In this Letter, we describe an 
algorithm for simulating quantum dynamics on small, noisy, universal
quantum computers using the idea of {\em restarting} the dynamics after each
time step. 
In order to advance a wavefunction from time $t$ to $t+\Delta t$,
a Trotter time step is first carried out on the 
quantum hardware. Rather than continuing the propagation, a variational
ansatz is fit to the result and this ansatz is used for the next time step.
By including knowledge of the form of the solution
directly into the ansatz, such as the symmetries it must obey, this procedure
provides a significant resilience to decoherence, allowing for much longer
propagation times with high fidelity. We demonstrate the method on multiple
instantiations of the Aubry-Andr\'e model with interactions at half-filling,
using simulations of noisy 
quantum computers with various levels of decoherence. Our restarted quantum
dynamics (RQD) algorithm is able to maintain a high-degree of fidelity ($>$0.9) for
hundreds of time steps after the standard Trotter approach has lost all
coherence and reached fidelities of zero. We show that, when using RQD, even a
near-term quantum computer will be able to prepare interesting, many-body
localized states.

\begin{figure}
  \centering
  \sidesubfloat[]{\includegraphics[width=0.9\columnwidth]{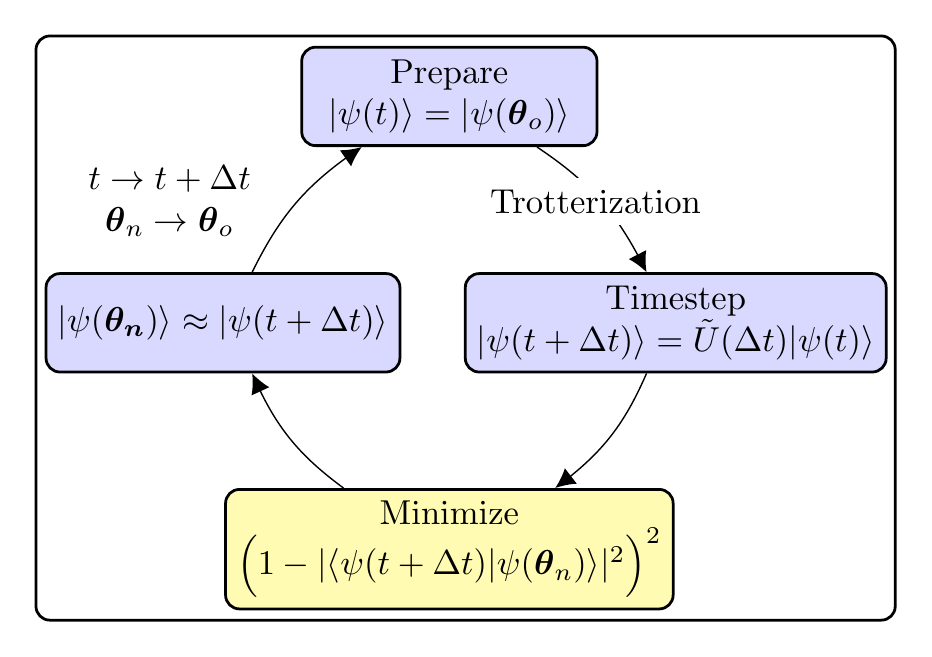}}\\
  \sidesubfloat[]{\includegraphics[width=0.9\columnwidth]{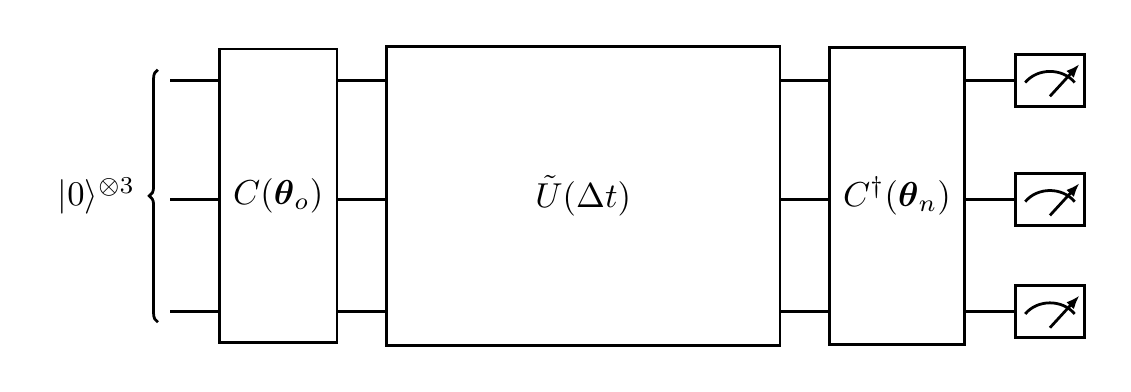}}\\
  \sidesubfloat[]{\includegraphics[width=0.9\columnwidth]{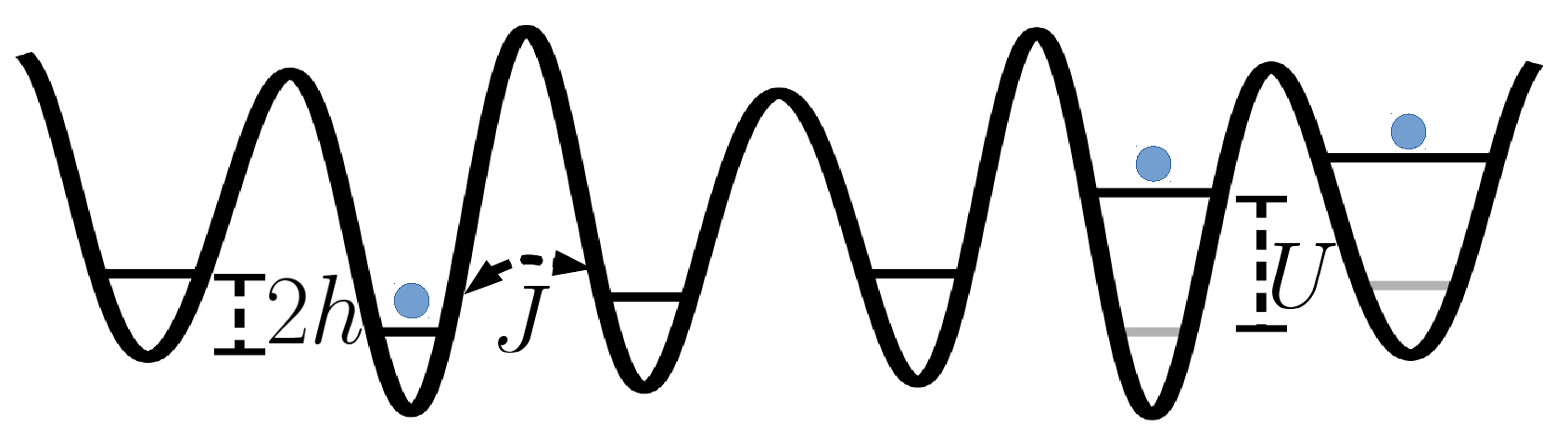}}

  \caption{
    (a) Diagram of restarted quantum dynamics (RQD) method. Blue shading
    represents steps taken on the quantum computer; yellow shading represent
    steps taken on the classical computer. (b) An example of a
    circuit used in the RQD method to estimate the
    fidelity between the time-stepped wavefunction and the new wavefunction
    ansatz. (c)
    Diagram of Aubry-Andr\'e model with interactions.  } \label{rqd_diagram} 
\end{figure}

\textit{Method.}
Let $H$ be some Hamiltonian and $|\psi(t)\rangle$ be the true wavefunction at some
time $t$. We seek the dynamics of the wavefunction given initial condition
$|\psi(0)\rangle$.
Rather than propagate all time steps on the quantum computer directly, we
propose to instead ``restart'' 
the dynamics after each time step by optimizing a wavefunction to approximate
the time-stepped wavefunction. We make use of a parameterizable circuit,
$C(\boldsymbol{\theta})$, giving a wavefunction 
\begin{equation} \label{ansatz}
  |\psi(\boldsymbol{\theta})\rangle = C(\boldsymbol{\theta}) | 0 \rangle,
\end{equation}
where $\boldsymbol{\theta}$ is a set of parameters, such as rotation angles, of
a circuit and $|0\rangle$ is 
the initial all zeros state of the quantum computer.
We take at least one time step directly on the quantum
computer via, e.g., a Trotter-like procedure. We then optimize the parameters of
a new wavefunction, $|\psi(\boldsymbol{\theta}_n)\rangle$, to the time-stepped
wavefunction, solving the following optimization problem
\begin{equation}\label{opt_trotter_swap}
  \min\limits_{\boldsymbol{\theta}_n} \Big(1 - |\langle \psi(\boldsymbol{\theta}_n) | \psi(t+\Delta t) \rangle|^2\Big)^2,
\end{equation}
where $|\psi(t+\Delta t) \rangle$ is the time-stepped wavefunction.
The RQD method is shown schematically in
Fig.~\ref{rqd_diagram}(a). 

The proposed algorithm can be summarized as three basic steps: 1) prepare the wavefunction 
$|\psi(t)\rangle$, which may itself be a circuit with a set of optimal
parameters found in a previous iteration; 2) take a single Trotter time step
directly on the quantum computer; 3) optimize the
fidelity of the time-stepped wavefunction with the variational wavefunction for
parameter set $\boldsymbol{\theta}_n$.
The
fidelity between two wavefunctions, $|\langle \psi(\boldsymbol{\theta}) | \psi(t+\Delta t)
\rangle|^2$, can be calculated in linear time with a generalized \texttt{SWAP}
test between two registers~\cite{cincio2018learning} or by appending the
reverse of variational circuit after the time
step and measuring to population of the all zero
state~\cite{havlivcek2019supervised}, as shown in Fig.~\ref{rqd_diagram}(b).  
After an optimal set of parameters is found, the variational wavefunction
approximates the time-stepped wavefunction
$|\psi(\boldsymbol{\theta}_n)\rangle \approx |\psi(t+\Delta t) \rangle$.
In the absence of noise and with a fully flexible, optimized wavefunction
ansatz this procedure would exactly reproduce the Trotter trajectory. With
noise, but keeping the same fully flexible ansatz, this procedure would 
fit a wavefunction including the undesirable effects of the decoherence. For example, if the
dynamics of the Hamiltonian were to put a two-qubit system into the $|01\rangle$ state,
but there was strong amplitude-damping noise in the system, the RQD 
method would optimize the wavefunction to give the decohered
state with a significant population in the $|00\rangle$ state. However, if the
ansatz is constructed to only allow certain states, the RQD 
method can provide noise-resiliency. If the ansatz were only capable of
preparing superpositions of $|01\rangle$ and 
$|10\rangle$, the optimized ansatz wavefunction would have no significant
elements of the $|00\rangle$ state, effectively correcting the amplitude-damping
noise. We emphasize a symmetry-preserving circuit ansatz allows for much longer
simulation times and constitutes one of the primary contributions of this
Letter. By placing more restrictions on the wavefunction, such as ensuring the
wavefunction preserves the same symmetries as the Hamiltonian, more
noise-resilience can be added to the algorithm. As long as the quantum computer
has enough coherence time 
to effectively prepare the ansatz twice and take a single Trotter time step (see
Fig.~\ref{rqd_diagram}(b)),
this method can be `restarted' many times.
To demonstrate this, we use the
Aubry-Andr\'e model with interactions at
half-filling~\cite{aubry1980analyticity}, a prototypical 
Hamiltonian which can demonstrate many-body
localization~\cite{iyer2013many,khemani2017two}.   

\textit{Aubry-Andr\'e Model.}
Here, we study the
one-dimensional spinless 
Aubry-Andr\'e model with interactions, 
given by the Hamiltonian 
\begin{multline}
  H = -J \sum_{k=1}^N (a_k^\dagger a_{k+1} + a^\dagger_{k+1} a_k) + h
  \sum_{k=1}^N \cos (2 \pi \beta k + \phi) a^\dagger_k a_k
  \\+ U \sum_{k=1}^N a^\dagger_k a_k a^\dagger_{k+1}a_{k+1} ,
\end{multline}
where $N$ is the number of sites, $J$ is the hopping strength, $h$ is the
disorder strength, $\beta$ is an irrational number, $\phi$ is a phase offset,
$U$ is the interaction strength, and we impose periodic boundary conditions. 
The second term describes the disorder, which can be described by the
interaction of two lattices with a ratio of periodicities $\beta$ and phase
offset $\phi$~\cite{aubry1980analyticity}. For almost all irrational $\beta$,
the Aubry-Andr\'e model 
shows interesting localization effects~\cite{castro2019aubry}. This model is shown
schematically in Fig.~\ref{rqd_diagram}(c).
Methods using matrix product
states can efficiently evolve such systems for short times, but the entanglement
grows as the dynamics proceeds, due to the fact that
dynamics involves many excited, entangled states~\cite{schuch2008entropy},
increasing the necessary bond dimension~\cite{doggen2019many}.
Since quantum computers 
live within the full Hilbert space, they can naturally represent such
highly entangled superpositions. 

One example of an interesting dynamical quantity
is the breaking of ergodicity due to many-body localization, which has been
demonstrated experimentally for small numbers of
spins~\cite{schreiber2015observation}. Starting from an 
initial charge density wave where even sites are unoccupied and odd sites are
occupied, 
$|\psi(0)\rangle = |1,0,1,0,1,0...\rangle$, the evolution of the imbalance,
\begin{equation}
  I(t)  = \frac{N_e(t) - N_o(t)}{N_e(t) + N_o(t)},
\end{equation}
where $N_e(t)$ is the occupation of even sites and
$N_o(t)$ is the 
occupation of odd sites. 
A nonzero imbalance at long times implies that the
state is many-body localized~\cite{andraschko2014purification}.

\textit{Results.}
Using simulations of noisy quantum computers, we calculate the dynamics of many
instantiations of the Aubry-Andr\'e model with interactions, 
fixing $\beta$ = $\sqrt{2}$, $\frac{U}{J}=\frac{h}{J}=4$, and
choosing $\phi$ randomly from a uniform distribution in the range $[0,2\pi]$. We
compare three different 
strategies for calculating the dynamics: Trotter, where we apply the
Trotterized evolution circuit multiple times, and the RQD 
method with two different ansatzes. The first ansatz is a number-conserving
circuit which, for any value of the parameters, conserves the total particle
number~\cite{gard2019efficient}. Since the Aubry-Andr\'e model conserves total
particle number, this 
number-conserving ansatz directly builds in an important symmetry of the
Hamiltonian. The second ansatz is an idealized, `oracle' ansatz of the form
$C(\theta) = \exp(-i \theta H)$, which is the true propagator with the time as
the single parameter and has all of the possible symmetries of the Hamiltonian.
We use a time step of $\Delta t = 0.04$ in all instances. We minimize
Eq.~\eqref{opt_trotter_swap}, 
which maximizes the fidelity, using 
L-BFGS~\cite{liu1989limited} with numerical gradients up to a tolerance of
$10^{-12}$ or for 
a maximum of 80 iterations. Analytical gradients could also be calculated
directly on the quantum computer by various means, such as shifting the rotation
angles by specific values~\cite{schuld2019evaluating}.

We simulate the evaluations of these circuits on a noisy quantum computer at
various noise rates using the density 
matrix master equation formalism~\cite{otten-pra-2016,otten-prb-2015}, as
implemented in the high-performance 
quantum dynamics simulator, QuaC~\cite{QuaC:17}. Noise is modeled using both
environmental amplitude-damping ($T_1$) and pure dephasing ($T_2^*$) noise.
Gates are treated as perfect, unitary operations which happen at a time $t$ and
are followed by a wait time consistent with the gate times of IBM's
superconducting qubit quantum computers (100-1000ns)~\cite{kandala-nature-2017}. We
treat the oracle ansatz as a single gate which takes the same amount of time as
the whole of the number-conserving ansatz, namely 0.026ms. Due to variance in the
effectiveness of the compilation and scheduling, the evolution
circuits take between 0.102ms and 0.245ms per time step. Since we use the
density matrix formalism, we have no stochastic sampling errors. Additional
details about the generation of the propagation circuits and the noisy quantum
computer simulations can be found in the appendix.

\begin{figure}
  \centering
  \includegraphics[width=\columnwidth]{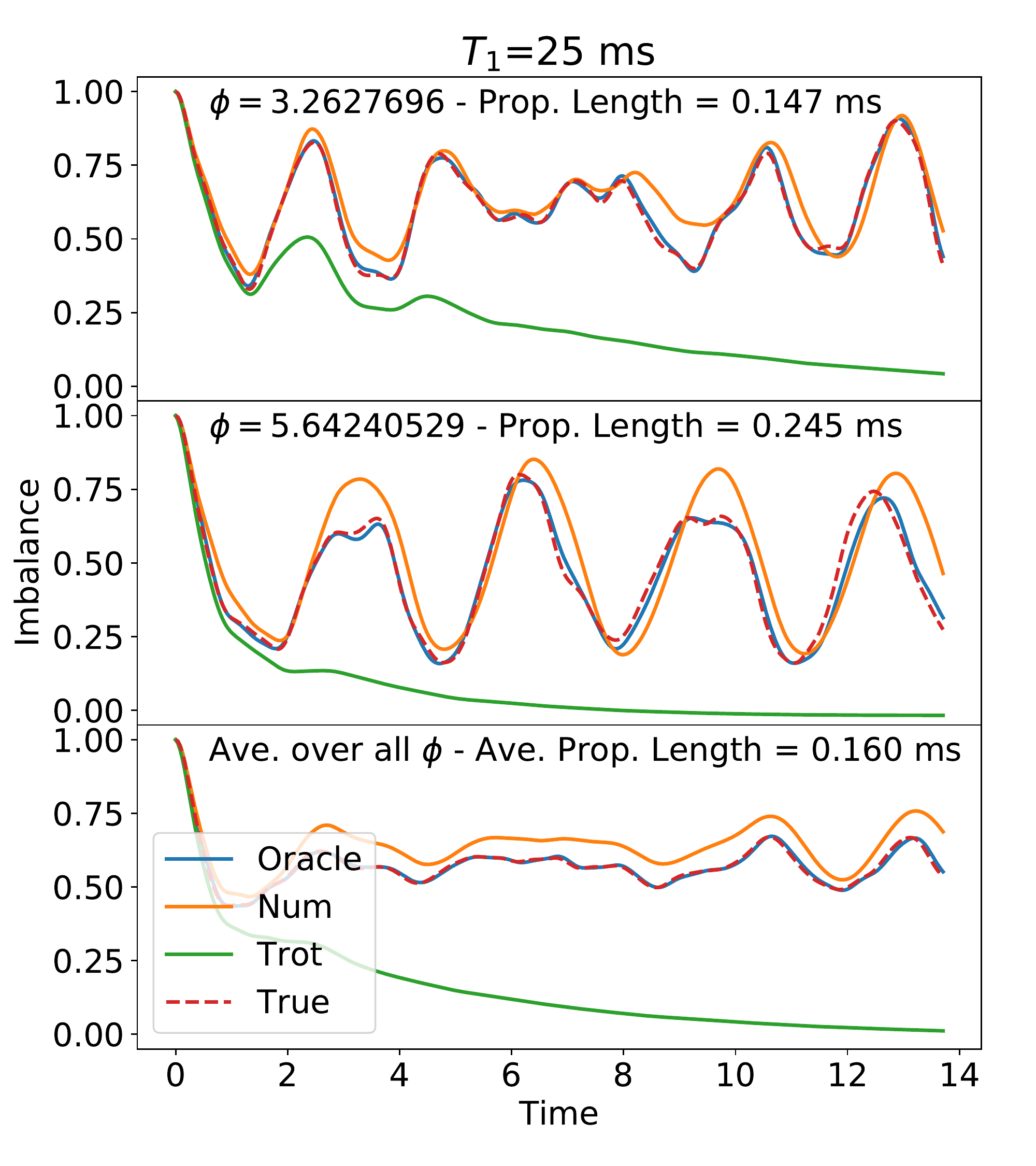}

  \caption{
    Examples of imbalance calculated with RQD and Trotter. RQD maintains both
    qualitative and quantitative agreement with the true dynamics, whereas the
    standard Trotter procedure quickly decays to zero.
  } \label{rqd_imbalance}
\end{figure}

Fig.~\ref{rqd_imbalance} shows the calculated imbalance for two specific
instantiations of $\phi$, as well as the average over all 16 calculations on a
simulated noisy quantum computer with $T_1=T_2^*=25$ms.
The average propagation circuit over
all $\phi$ values compiles to 0.160ms per time step. Roughly, this implies that
a standard Trotter procedure could take between 10 and 20 time steps before the total
circuit time exceeds a tenth of the coherence time of our simulated quantum
computer and the noise effects would begin to greatly affect the output.
This can be seen clearly in Fig.~\ref{rqd_imbalance}. In
the shorter circuit, with $\phi=3.267696$ and a propagation circuit length of
0.147ms per time step, the standard Trotter procedure is able to weakly capture
one 
oscillation at around $t= 2$, after 50 time steps and a total
propagation time of around 8ms. In the longer circuit, with
$\phi=5.64240529$ and a propagation circuit length of 0.245ms, the standard
Trotter procedure decays much faster, only capturing a hint of the first
oscillation.

The RQD method, on the other hand,
quantitatively captures the dynamics for the whole time window calculated. 
For the shorter circuit of $\phi=3.267696$, the number-conserving ansatz is able
to capture most of broad oscillations and even many of the smaller oscillations,
whereas in the longer circuit of $\phi=5.64240529$, it captures only the broad
oscillations. The oracle ansatz, on the other hand, captures essentially all of
the features. The
oracle ansatz encodes all of the possible symmetries of the wavefunction,
whereas the number-conserving ansatz only has the single, but important,
particle number symmetry. This allows the oracle ansatz to effectively recover
from more errors than the number-conserving ansatz. The oracle ansatz is also
much simpler to optimize, having only a single parameter, compared to the 38
parameters of the number-conserving ansatz, allowing it to reliably converge to
our optimization tolerance of $10^{-12}$.
The number-conserving ansatz almost always reached the iteration limit before
converging.

The lower panel of Fig.~\ref{rqd_imbalance} shows the average imbalance over all
16 $\phi$ values. The parameters of the Aubry-Andr\'e model we chose
($\frac{U}{J}=\frac{h}{J}=4$) are beyond the critical disorder strength
($\frac{h}{J}=2$) for localization. As such, the true dynamics show a
significant nonzero imbalance at long times. For both ansatzes, the RQD 
method agrees well with the true dynamics, showing the many-body
localization effect. The Trotter dynamics, on the other hand, only reproduces
the correct imbalance before $t=1$. In the longer times,
the imbalance has decayed to 0, and the many-body localization effect is not
observable. 

\begin{figure}
  \centering
  \includegraphics[width=\columnwidth]{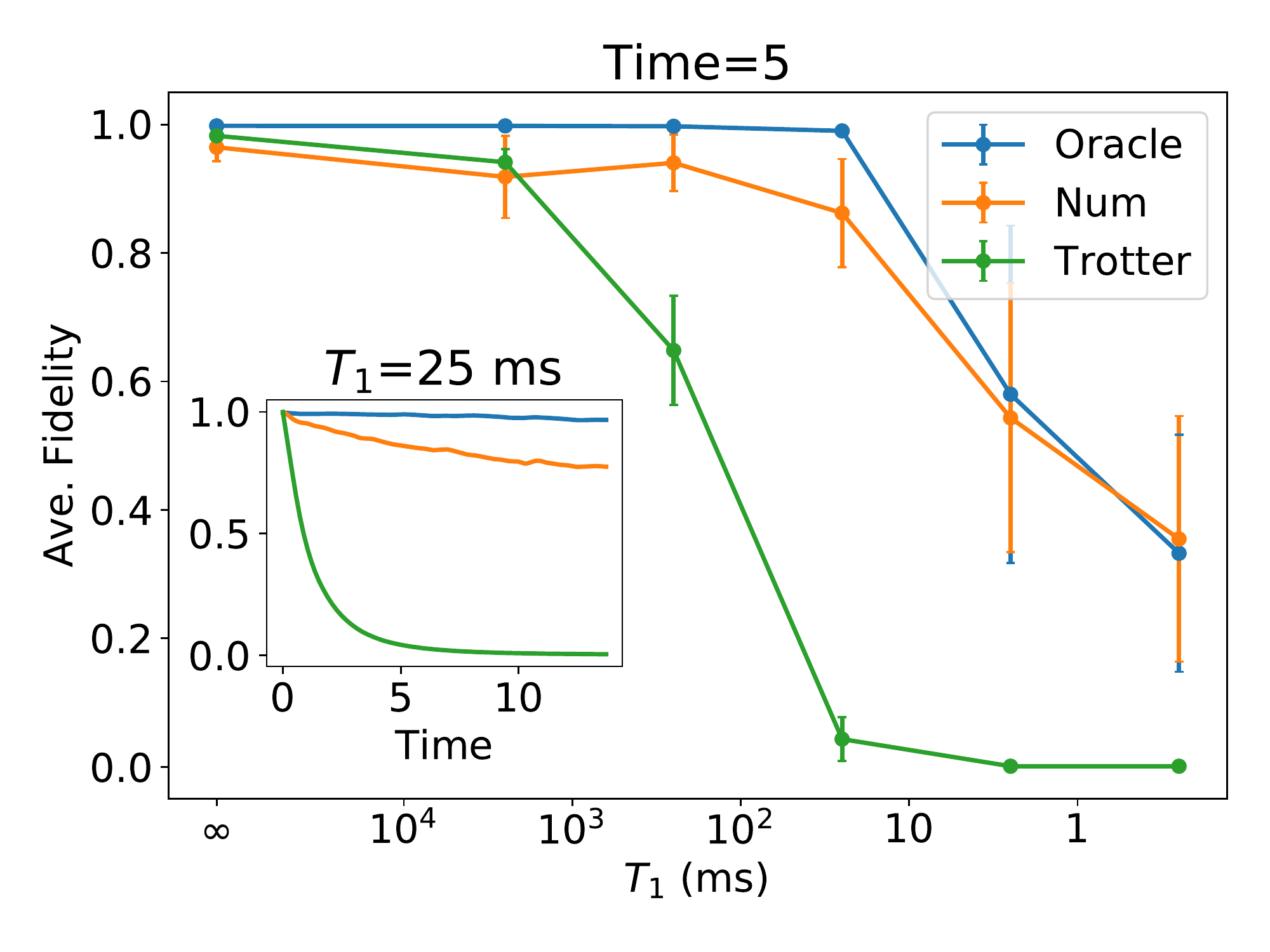}
  \caption{
    Comparison of fidelity with RQD and Trotter, averaged over sixteen
    instantiations of the Aubry-Andr\'e model with interactions. 
  } \label{rqd_fidelity}
\end{figure}

Figure~\ref{rqd_fidelity} shows the average fidelity over all 16 values of
$\phi$ at different $T_1=T_2^*$ coherence times for the time slice $t=5$. At
very long coherence times  
(very low noise rates), including the noise-free case, the standard Trotter
procedure performs better than RQD with the
number-conserving ansatz, but worse than the oracle ansatz. The oracle ansatz
even recovers from Trotter errors, as it can only prepare states that are
part of the true dynamics. The number-conserving ansatz, on the other hand, is a
more flexible wavefunction that can, in principle, fit the Trotter wavefunction,
even with Trotter errors. It performs worse than Trotter for long coherence
times due to the optimization failing to converge to a small enough tolerance.
As the coherence times decrease (and the noise rates increase), the fidelity of the
standard Trotter procedure drops quickly, going to nearly zero at a coherence time of only
25ms. The oracle and number-conserving ansatzes both maintain high fidelity at
25ms. In fact, the fidelity of RQD with either ansatz is
competitive with the standard Trotter procedure at coherence times two orders
of magnitude smaller.

The inset of Fig.~\ref{rqd_fidelity} shows the average fidelity over all 16
values of $\phi$ for the coherence time $T_1=T_2^*=25$ms throughout the
calculation of the model dynamics. The average fidelity of the standard Trotter
procedure quickly drops off as more time steps are taken due to the increasing
circuit depth. The RQD method, with either ansatz,
maintains a high fidelity.
The oracle ansatz maintains a fidelity of nearly 1,
whereas the number-conserving ansatz decays slowly. The oracle ansatz encodes
far more symmetries than the number-conserving ansatz, making it inherently more
robust to more types of errors. For example, pure dephasing errors, which are
parameterized by $T_2^*$, can maintain particle number but still cause the
fidelity to decrease. The number-conserving ansatz is, therefore, not robust to
these errors and they gradually build up as the calculation proceeds. This is in
addition to errors from optimization, which also gradually build up. When the
optimization terminates due to reaching the iteration limit, the next time step
comes from a wavefunction which approximates the time-stepped wavefunction worse
than otherwise possible, leading to a build up of error as more time steps are
taken. One way to mitigate the build up of such optimization errors as quantum
computers with longer coherence times are built is to take multiple Trotter time
steps before the restart procedure. Utilizing recent advancements in error
mitigation techniques, such as error
extrapolation~\cite{otten2019recovering,otten2019accounting,kandala2019error,temme2017error,li2017efficient}, 
could help allow multiple Trotter time steps to be taken. 

The key overhead of the RQD method is
optimization. For realistic ansatzes, such as the number-conserving ansatz used
here, the number of parameters will generally be large and the number of
iterations needed for convergence will also become large. Even though each
function evaluation involves measuring only a single observable, the fidelity,
there is need for algorithmic 
improvement in the optimization subroutine to ensure the method remains
practical for large systems. Recent advancements in optimization for other hybrid
quantum-classical methods~\cite{verdon2019learning,shaydulin2019multistart} can be applied to 
restarted quantum dynamics to alleviate this overhead. Furthermore, efficient
ansatz design, which is problem specific, could restrict the number of free
parameters, providing more robustness to noise and additional ease of
optimization.

\textit{Conclusion.}
We described and demonstrated an algorithm for carrying out dynamics
calculations on lossy, near-term quantum computers using the idea of
``restarting'' the dynamics. This restarting procedure involves approximating
the time-stepped wavefunction with some variational ansatz which is optimized to
give the result of the time step. By encoding known symmetries of the true
wavefunction into the ansatz, the RQD method is able to
mitigate the effects of noise during the propagation at the additional cost of
optimization. Careful ansatz design which includes \textit{a priori} knowledge
is necessary to provide the noise mitigation. A completely flexible ansatz would
learn the noise effects, along with the dynamics. We demonstrated that RQD 
with only number-conservation symmetry in the ansatz was able
to greatly extend the length of calculation of the Aubry-Andr\'e model with
interactions that could be computed on a 
simulation of a noisy quantum computer. At a coherence time of $T_1=T_2^*=25$ms,
the RQD approach was able to take hundreds of time steps
beyond where the standard Trotter procedure began to fail.
With additional
symmetries, RQD performs even better. 
 Restarted quantum dynamics is a promising
algorithm that could pave the way for quantum dynamics calculations on quantum
computers which take many more time steps than the coherence time would naively
allow. 

\begin{acknowledgments}
  This work was performed at the Center for Nanoscale Materials, a U.S. Department
  of Energy Office of Science User Facility, and supported by the U.S. Department
  of Energy, Office of Science, under Contract No. DE-AC02-06CH11357. We
  gratefully acknowledge the computing resources provided on Bebop,
  a high-performance computing cluster operated by the Laboratory Computing
  Resource Center at Argonne National Laboratory. 
\end{acknowledgments}

\appendix

\section{Propagation Circuit Generation}
\begin{table}
  \begin{tabular}{ cccc }
    \hline
    $\phi$ & Total Time (ms) & Layers & Gates\\ 
    \hline
    1.93146731  & 0.1158  & 163 & 187\\
    5.64240529  & 0.2452  & 312 & 347\\
    1.57973617  & 0.1602  & 212 & 246\\
    0.08769829  & 0.2264  & 291 & 324\\
    4.42879993  & 0.1642  & 219 & 251\\
    1.59366522  & 0.1822  & 238 & 270\\
    1.69972758  & 0.1494  & 203 & 230 \\
    3.26279226  & 0.1472  & 197 & 223 \\
    6.09740422  & 0.2302  & 294 & 329 \\
    3.34460202  & 0.2632  & 335 & 372 \\
    3.26276960  & 0.1026  & 148 & 170 \\
    4.52159699  & 0.2102  & 274 & 308 \\
    2.94545992  & 0.1522  & 203 & 234 \\
    4.71502552  & 0.2184  & 283 & 314 \\
    1.08255072  & 0.2122  & 274 & 306 \\
    4.85940981  & 0.1022  & 146 & 172 \\
    \hline
  \end{tabular}
  \caption{Values of $\phi$ and the time, number of layers, and total number of
    gates for the corresponding evolution circuits.}\label{circuit_table}
\end{table}

We first generate the
fermionic Hamiltonian for a given $\phi$ and apply the Jordan-Wigner
transform~\cite{whitfield-molphys-2011}, using
OpenFermion~\cite{mcclean-arxiv-2017}, to obtain the 
qubit Hamiltonian. We then 
generate the evolution operator using a first-order Trotter procedure~\cite{berry2007efficient}
with time step $\Delta t = 0.04$, and compile the resulting
circuit for a fully-connected quantum computer using Qiskit~\cite{qiskit}. Due to
differences in the compilation and scheduling procedures (see below), the
resulting circuits 
end up with different final numbers of layers. The sixteen values of $\phi$ and
the resulting circuit lengths are shown in Table~\ref{circuit_table}.

\section{Scheduling Algorithm}
We use a simple, greedy algorithm to schedule the gates into layers where all
gates can be applied in parallel. The algorithm begins with a list of gates
which are guaranteed to be in correct sequential order. We also initialize the
current layer counter for each qubit to zero. In sequence, we check
the qubits to which each gate is applied. We assign the current gate to the
maximum of the current layers for the constituent qubits. We then set the
current layer counter of the constituent qubits to the maximum of the current
layers plus one. This process is done until all gates have been assigned layers.
The layers are then applied in sequence, with layers potentially being comprised
of multiple, parallel gates. 

\section{Simulation of Noisy Quantum Computer}
We use the Lindblad master equation to simulate a noisy quantum computer. The
Lindblad master equation for a general system is
\begin{equation}
  \frac{\mathrm{d} \hat{\rho}}{\mathrm{d} t}
  = -\frac{i}{\hbar} [ \hat{H}, \hat{\rho} ] +
  L(\hat{\rho}),
  \label{master_eqn}
\end{equation}
where $\hat{H}$ is the system Hamiltonian and $L(\hat{\rho})$ is the Lindblad
superoperator describing decoherence effects. For our system, we use no direct
Hamiltonian $H$ and we simulate amplitude-damping ($T_1$) and pure-dephasing ($T_2^*$)
decoherence sources,
\begin{multline}\label{5_lindblad_q}
  L_i(\hat{\rho}) =
  - \frac{1}{2 T_1}
  (\hat{\sigma}_i^{\dagger} \hat{\sigma}_i \hat{\rho}
  + \hat{\rho} \hat{\sigma}_i^{\dagger} \hat{\sigma}_i
  - 2 \hat{\sigma}_i \hat{\rho} \hat{\sigma}_i^{\dagger}) \\
  - \frac{1}{T_2^*}
  (\hat{\sigma}_i^{\dagger} \hat{\sigma}_i \hat{\rho}
  + \hat{\rho} \hat{\sigma}_i^{\dagger} \hat{\sigma}_i
  - 2 \hat{\sigma}_i^{\dagger}
  \hat{\sigma}_i \hat{\rho} \hat{\sigma}_i^{\dagger} \hat{\sigma}_i),
\end{multline}
where $L_i(\hat{\rho})$ is the Lindblad superoperator for qubit $i$,
$\hat{\sigma}_i$ is the annihilation operator for qubit $i$, and all qubits have
the same decoherence times, $T_1$ and $T_2^*$.

\section{Number Conserving Ansatz}
\begin{figure}
  \centering
  \includegraphics[width=\columnwidth]{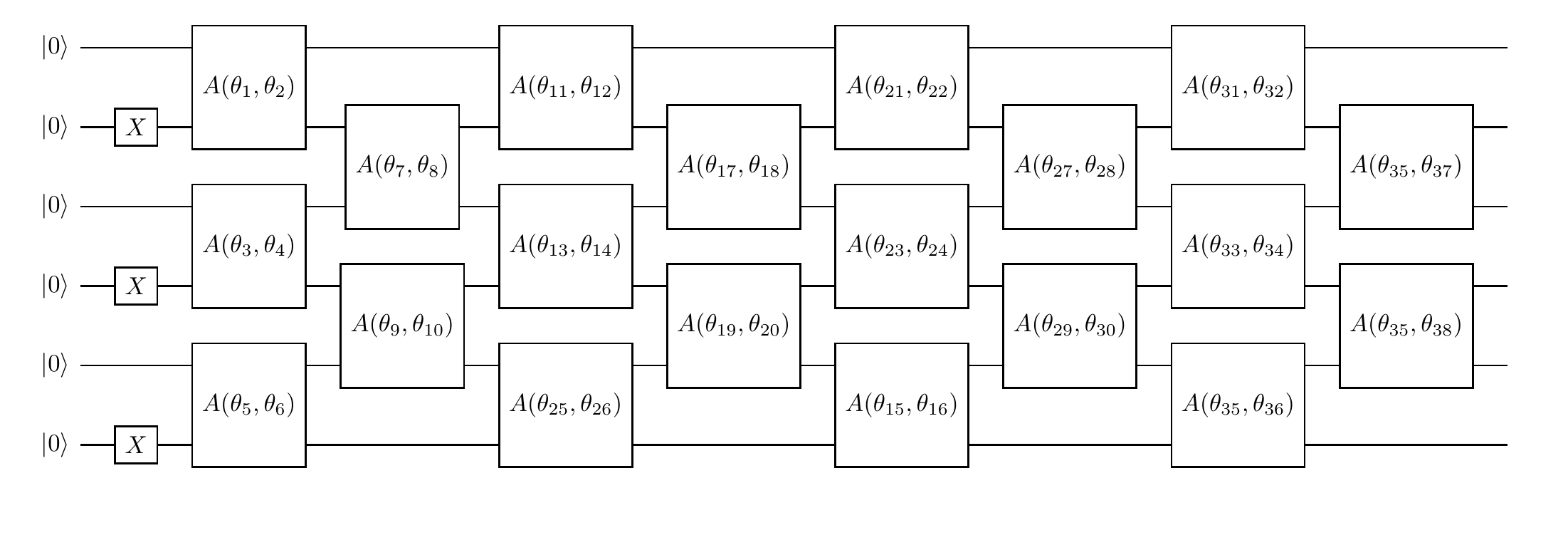}
  \caption{
    Number-conserving ansatz for three particles in six sites. 
  } \label{number_ansatz}
\end{figure}

\begin{figure}
  \centering
  \includegraphics[width=\columnwidth]{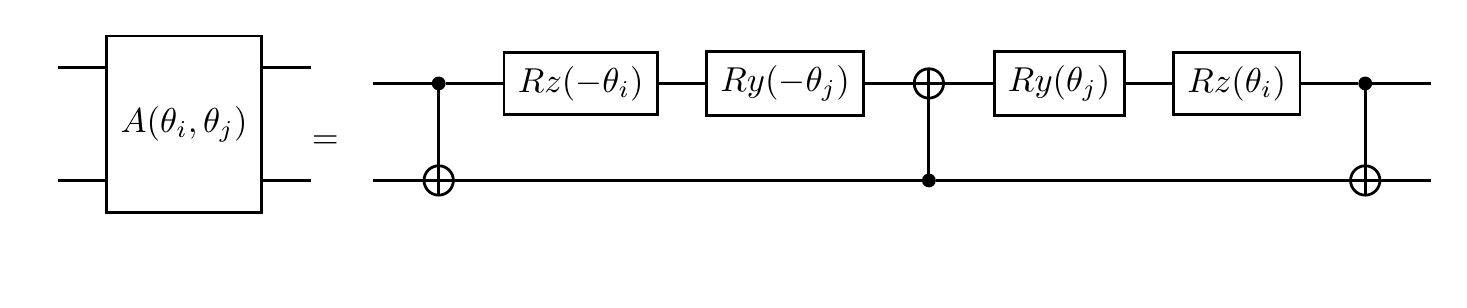}
  \caption{
    `A' gate used in number-conserving ansatz.
  } \label{number_ansatz_A}
\end{figure}

Figure~\ref{number_ansatz} shows the explicit number-conserving ansatz used in
the main text, using the `A' gate of figure~\ref{number_ansatz_A}. This ansatz
conserves three particles within six sites. It was generated according to the
method of~[\onlinecite{gard2019efficient}]. 

\onecolumngrid

\twocolumngrid

\bibliography{restarted_dynamics}

%

\end{document}